\documentclass{article}\usepackage{jkas2}

\runningauthor{CHO}
\runningtitle{Incompressible MHD Turbulence}

\begin{document}

\title{NUMERICAL SIMULATIONS OF INCOMPRESSIBLE MHD TURBULENCE} 

\author{JUNGYEON CHO}

\address{ 475 N. Charter St., 
   Department of Astronomy, Univ. of Wisconsin, Madison, WI53706, USA;
   cho@astro.wisc.edu}

\address{\normalsize{\it (Received ???. ??, 2001; Accepted ???. ??,2001)}}

\abstract{
       The study of incompressible magnetohydrodynamic (MHD) 
       turbulence gives useful insights
       on many astrophysical problems.
       We describe a pseudo-spectral MHD code suitable for the study of
       incompressible turbulence.
       We review our recent works on
       direct three-dimensional numerical simulations 
       for MHD turbulence in a periodic box.
       In those works, 
       we use a pseudo-spectral code to solve the incompressible MHD 
       equations.
       We first discuss the structure and properties of turbulence 
       as functions of scale.
       The results are consistent with the scaling law recently proposed
       by Goldreich \& Sridhar. 
       The scaling law is based on the concept of scale-dependent isotropy: 
       smaller eddies are more elongated than larger ones along magnetic
       field lines.
       This scaling law substantially changes our views on MHD turbulence. 
       For example, as noted by Lazarian \& Vishniac, the scaling law
       can provide a fast reconnection rate. 
       We further discuss how the study of incompressible MHD turbulence can
       help us to understand physical processes in interstellar medium (ISM)
       by considering imbalanced cascade and viscous damped turbulence.
}

\keywords{ISM:general-MHD-turbulence}

\maketitle

\section{INTRODUCTION}

Astrophysical flows are complicated and dynamic.
{}For example, molecular clouds are clumpy over
a vast range of scales and observations of 
velocity spectral line broadening indicate that
dynamic pressure ($\rho\,{\bf
v}^2$/2) usually dominates the thermal pressure ($nkT$).
{}Furthermore, velocity spectral lines show multiple components.
These facts suggest that cloud internal velocity
possesses disordered components.
In molecular clouds, the characteristic Reynolds number 
is estimated to be larger than $\sim 10^{10}$, which is consistent with
the turbulent state in the clouds.
In general, when the Reynolds number ($\equiv LV/\nu$; L=characteristic
size of the system, V=velocity dispersion, and $\nu$=viscosity) is larger than
an order of 10 - 100, the system becomes unstable and turbulent.

{}In laboratory systems, fluids are usually incompressible and
unmagnetized.
Kolmogorov phenomenology provides an excellent insight
for such systems.
Suppose that we `disturb' the fluid at a scale $L$.
We call this scale as {\it energy injection scale} or 
{\it largest energy containing eddy scale}.
Then the energy injected to the scale $L$ cascades to progressively
smaller and
smaller scales.
Ultimately, the energy will reach the molecular dissipation scale $l_d$
and it will be lost there.
The scales between $L$ and $l_d$ is called the {\it inertial range}.
Suppose a scale $l$ lies in the inertial range.
Let the characteristic velocity associated with the scale be $v_l$.
Kolmogorov theory states that the kinetic energy ($v_l^2$) is
transferred to one-level smaller scale within one eddy turnover
time ($l/v_l$). It is natural that the cascade rate ($v_l^2/(l/v_l)$)
be scale-independent.
Therefore, we have
\begin{equation}
 v_l \propto l^{1/3}. \label{eq1}
\end{equation}
(One-dimensional) Energy spectrum $E(k)$ is one of the most important 
quantities in turbulence
theories.
Note that $E(k) dk$ is the amount of energy between the wavenumber $k$
and $k + dk$.
When $E(k)$ follows a power law, $kE(k)$ is the energy {\it near} the 
wavenumber $k\sim 1/l$.
Since $v_l^2$ represents a similar energy, we have $v_l^2 \approx kE(k)$.
Therefore, equation (\ref{eq1}) becomes
\begin{equation}
  E(k) \propto k^{-5/3}.
\end{equation}
This is the well-known Kolmogorov spectrum.
Astrophysical fluids are compressible and magnetized.
Nevertheless, many astrophysical quantities follow 
this Kolmogorov spectrum.
It is important to note that Kolmogorov theory predicts: 1.
the velocity dispersion ($\sim v_l$) decreases with scales as 
$v_l \propto l^{1/3}$; 2. the slope of the energy spectrum is $-5/3$.

Earlier works on ISM focused on the velocity dispersions
and derived a qualitative agreement with the Kolmogorov theory.
Larson (1981) and Myers (1983) found that
velocity dispersion decreases as the size of the clumps
decreases, which
may support turbulent model of ISM.
Scalo (1984) showed that two-point velocity correlation increases
as the separation decreases, which is also qualitatively consistent
with turbulent model.
All the above mentioned studies are for scales larger than or similar to
$\sim$ 1 pc.

On the other hand, other studies address the issue of power spectra. 
Interstellar scintillation observations
indicate electron density spectrum follows a power law over
7 decades of length scales (see Armstrong et al. 1995).
The slope of the spectrum is very close to $-5/3$ for
$10^6 m$ - $10^{14} m$. The spectrum is somewhat uncertain
{}for scales larger than $10^{14} m$ ($\sim 0.01pc$).
However, recently, Lazarian \& Pogosyan (2000) and 
Starnimirovic \& Lazarian (2001) showed the
existence of the $-5/3$ velocity power spectrum over pc-scales in HI.
Solar-wind observations provide {\it in-situ} measurements of the
power spectrum of magnetic fluctuations.
Leamon et al. (1998) obtained a slope of $-1.7$.

Therefore it is evident that turbulence is prevalent in astrophysical
{}fluids.
Understanding the nature of such turbulence is important
because it affects many astrophysical processes.
{}For example, star formation rate depends on the 
turbulence decay time-scale.
Heat and cosmic ray transport also depend on the statistics of
turbulence.
Recently anisotropic MHD Turbulence has also been evoked to 
provide fast magnetic reconnection rate (Lazarian \& Vishniac 1999).

One approach to studying the ISM is to perform time-dependent numerical
simulations to model the ISM, including as many of the
interacting phenomena as practical. Of course, the physics
included in such models must necessarily be highly
simplified, and it is difficult to determine which features
of the final model result from which physical assumptions (or
initial conditions). Our approach is to use simplified
numerical simulations to study the influences of various
physical phenomena in isolation. We want to obtain a
physical feeling for the general effects that each
phenomenon has on the nature of the ISM. 
Here we review our previous works on
incompressible MHD turbulence.
It is obvious
that the real ISM is compressible, but we want to separate
the effects of magnetic turbulence from those involving
compression. In later works we will
include compression for comparison with the present models,
thereby isolating its importance directly.

As we said earlier, we will numerically 
study incompressible MHD turbulence.
We believe Fourier spectral method is one of
the best numerical methods for such calculations.
In II, we  show how the spectral method works.
In III, we introduce our recent works done with the spectral
code: anisotropy, decay speed of MHD turbulence, and magnetic structures below
the viscous cut-off.
In IV, we give conclusion.

\section{NUMERICAL METHOD}

\subsection{PSEUDO-SPECTRAL METHOD}

{}For simplicity, let us first consider incompressible {\it unmagnetized}
fluid.  Then, the governing equation is the Navier-Stocks equation:
\begin{equation}
  \frac{\partial {\bf v} }{\partial t} = -\nabla \cdot ({\bf v}{\bf v})
               + \nu \nabla^{2} {\bf v} + \nabla P ,
\end{equation}
where ${\bf v}$ is velocity, $\nu$ is viscosity, and $P$ is gas pressure.
Since we are considering an incompressible fluid,
we can assume that $\rho=1$.

The Fourier spectral method is highly suitable for incompressible flows.
Simply put, the idea of the method is simple: we transform all physical
space variables into Fourier (or wave-vector) space and
solve the differential equation there.
Hereafter, we assume that we are solving the equation on a uniform
rectangular grid.
Some advantage of the method is:

\noindent
1. It shows fast convergence for smooth functions. \newline
2. It is easy to implement divergence-free conditions.\newline
3. It is easy to implement periodic boundary condition. \newline
4. Fourier space analyses (e.g. calculation of energy   \newline   
   spectrum) are natural consequences.

We note that the gradient operator becomes $i{\bf k}$ and
that multiplication becomes convolution sum in Fourier space.
Here $i=\sqrt{-1}$.
Using these facts, we obtain the Navier-Stocks equation 
in Fourier space:
\begin{equation}
\frac{ \partial \hat{\bf v}({\bf k}) }{\partial t} = -i{\bf k}\cdot 
    (\widehat{ \bf vv })({\bf k}) -i\nu k^2 \hat{\bf v}({\bf k})
    +i{\bf k} \hat{P}({\bf k}),
\end{equation}
where
\begin{equation}
(\widehat{ \bf vv })({\bf k}) \equiv \int_{ {\bf p}+{\bf q} ={\bf k} }
 d{\bf k} ~\hat{\bf v}({\bf p}) \hat{\bf v}({\bf q}).
\end{equation}
The incompressibility condition  becomes
\begin{equation}
   i{\bf k}\cdot \hat{\bf v}({\bf k}) = 0,
\end{equation}
which means that
the velocity vector stays perpendicular to the wave vector ${\bf k}$.
Since the pressure term in equation (4) is parallel to ${\bf k}$, 
we can drop the
term. On the other hand, the nonlinear term (i.e the term
containing $\widehat{ \bf vv }$) may have
both parallel and perpendicular components.
Therefore, we need to remove the component parallel to ${\bf k}$ by
applying an appropriate projection operator.

The calculation of the nonlinear term is the most time consuming part of
the Fourier spectral methods.
{}For simplicity, let us consider the following 
one dimensional convolution
sum:
\begin{equation}
   \int_{p+q=k} dk ~\hat{f}(p)\hat{g}(q).
\end{equation}
Then the discrete representation of the convolution sum is
\begin{equation}
   \sum_{p+q=k} \hat{f}_p \hat{g}_q.
\end{equation}
The pseudo-spectral method approximates the convolution sum as
\begin{equation}
\sum_{p+q=k} \hat{f}_p \hat{g}_q \approx 
   \left[ F^{-1}\left[ F(\hat{f}) F(\hat{g}) \right] \right]_k,
\end{equation}
where $F$ and $F^{-1}$ denote forward/inverse Fourier transform.
The cost of such calculation is $O(N\log N)$, where $N$ is the 
number of grid points.

Once we obtain the convolution sum, we can use
any techniques for {\it ordinary} differential equations to 
solve the time evolution of the 
Navier-Stocks equation.

\subsection{ACTUAL SETUP}
In actual calculations, we have adopted a pseudospectral code and 
calculated the time evolution of incompressible magnetic turbulence
subject to a random driving force per unit mass:
\begin{equation}
\frac{\partial {\bf v} }{\partial t} = (\nabla \times {\bf v}) \times {\bf v}
      -(\nabla \times {\bf B})
        \times {\bf B} + \nu \nabla^{2} {\bf v} + {\bf f} + \nabla P' ,
        \label{veq}
\end{equation}
\begin{equation}
\frac{\partial {\bf B}}{\partial t}=
     \nabla \times ({\bf v} \times{\bf B}) + \eta \nabla^{2} {\bf B} ,
     \label{beq}
\end{equation}
\begin{equation}
      \nabla \cdot {\bf v} =\nabla \cdot {\bf B}= 0,
\end{equation}
where $\bf{f}$ is a random driving force,
$P'\equiv P/\rho + {\bf v}\cdot {\bf v}/2$, ${\bf v}$ is the velocity,
and ${\bf B}$ is magnetic field divided by $(4\pi \rho)^{1/2}$.
In this representation, ${\bf v}$ can be viewed as the velocity 
measured in units of the r.m.s. velocity, v,
of the system and ${\bf B}$ as the Alfven speed in the same units.
The time $t$ is in units of the large eddy turnover time ($\sim L/v$) and
the length in units of $L$, the inverse wavenumber of the fundamental
box mode.
In this system of units, the viscosity $\nu$ and magnetic diffusivity $\eta$
are the inverse of the kinetic and magnetic Reynolds numbers respectively.
The turbulence is driven by 21 random forcing components in Fourier space.
The peak of energy injection occurs at $k\approx 2.5 $.
The amplitudes of the forcing components are tuned to ensure $v \approx 1$
We use exactly the same forcing terms
for all simulations.
We use up to $256^3$ collocation points.
We use an integration factor technique for kinetic and magnetic dissipation terms
and a leap-frog method for nonlinear terms.
At $t=0$, the magnetic field has only its uniform component
and the velocity field is restricted to the range
$2\leq k \leq 4$ in wavevector space.

{}For some runs, we use hyperviscosity and hyperdiffusivity
{}for the dissipation terms.
The power of hyperviscosity
is set to either 2 or 8, so that the dissipation term in the above equation
is replaced with
\begin{equation}
 -\nu_h (\nabla^2)^h {\bf v},
\end{equation}
where $h=2$ or $8$.
A similar expression is used for the magnetic dissipation term.
In subsections III(a) and III(b), we use unit magnetic Prandtl
number ($\nu=\eta$) and the notation NY-$B_0$Z.
In subsection III(c), we use physical viscosity and physical- or
hyper-diffusion and the notation NXY-$B_0$Z.
Here N = 256, 144 refers to the number of grid points in each spatial
direction; X = P refers to physical viscosity;
Y = P, H2, H8 refers to the form of diffusion 
(physical-, the 2nd power  
hyper-, or the 8th power hyper-diffusion, respectively);
Z=0.5 or 1 refers to the strength of the external magnetic field in Alfven
velocity unit.

Diagnostics for our code can be found in Cho and Vishniac (2000b).

\section{PROPERTIES OF MHD TURBULENCE}
In this section, we discuss the properties of MHD turbulence
in the presence of a strong uniform background field.

\subsection{ANISOTROPY}

Historically hydrodynamic turbulence in an incompressible fluid was 
successfully
described by the eddy cascade (Kolmogorov 1941), 
but MHD turbulence was first modeled by 
wave turbulence (Iroshnikov 1963, Kraichnan 1965). 
This theory assumes isotropy of the energy cascade
in Fourier space.
However, the magnetic field defines a local symmetry axis since 
it is easy to mix
field lines in directions perpendicular to the
local ${\bf B}$ and much more difficult to bend them.
In a turbulent medium, the kinetic energy associated with large scale motions
is greater than that of small scales.
However, the strength of the local mean
magnetic field is almost the same on
all scales.
Therefore, it becomes relatively difficult to bend
magnetic field lines as we consider smaller scales,
leading to more pronounced anisotropy.
A self-consistent model of MHD turbulence which incorporates
this concept of scale dependent anisotropy was introduced
by Goldreich \& Sridhar (1995; hereinafter GS95).

The major predictions of the GS95 model are as follows.
{}First, anisotropy is scale dependent.
Roughly speaking, the semi-major axis ($1/k_{\|}$) and 
the semi-minor axis ($1/k_{\perp}$) of an eddy
satisfy the relation
$k_{\|}\propto k_{\perp}^{2/3}$,
which means smaller eddies are
relatively more elongated than larger ones.
Here $k_{\perp}$ and $k_{\|}$ are wave numbers perpendicular and parallel
to the background field.
Second, the model predicts that the one-dimensional energy spectrum 
is Kolmogorov-type if expressed in terms of perpendicular wavenumbers, i.e.
$E(k_{\perp})\propto k_{\perp}^{-5/3}$.

\begin{figure}[t]
\plotone{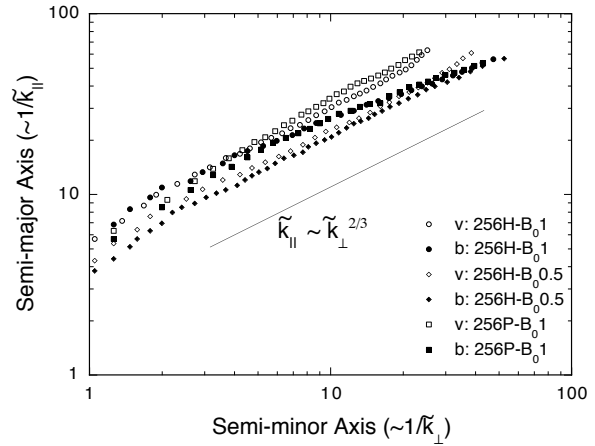}
\caption{  The relation between $k_{\|}$ and $k_{\perp}$. 
(Cho \& Vishniac 2000a).
 }
\end{figure}

Numerical simulations by Cho \& Vishniac (2000a) and 
Maron \& Goldreich (2001)
have mostly supported the GS95 model and helped to extend it. 
Both analyses stressed the point 
that scale dependent anisotropy can be measured only in 
local coordinate frames which are aligned with the locally averaged
magnetic field direction.
Cho \& Vishniac (2000a) calculated
the structure functions of velocity and magnetic field in the
local frames, and found that the contours of the structure functions do 
show scale dependent
anisotropy, consistent with the predictions of the GS95 model.
{}Fig 1. shows that the semi-major axis ($1/k_{\|}$)
is proportional to the 2/3 power of the semi-minor axis ($1/k_{\perp}$),
implying that $k_{\|}\propto k_{\perp}^{2/3}$.
One dimensional energy spectrum follows Kolmogorov spectrum:
$E(k)\propto k^{-5/3}$
 (see
Cho \& Vishniac 2000a).

\subsection{DECAY OF MHD TURBULENCE}

Turbulence plays a critical role in molecular cloud support and star 
formation and the issue of the time scale of turbulent decay is vital for
understanding these processes.

If MHD turbulence decays quickly then serious
problems face researchers attempting to explain important observational 
facts, i.e.  turbulent  motions seen within molecular clouds without
star formation (see Myers 1999) and rates of star formation (Mckee 2000).
Earlier studies attributed the rapid decay of turbulence to compressibility
effects (Mac Low 1999). Cho et al. (2001), as well as earlier ones  
(Cho \& Vishniac 2000a, Maron \& Goldreich 2001), 
shows that turbulence decays rapidly even
in the incompressible limit. This can be understood in the framework of
GS95 model where mixing motions perpendicular to magnetic field lines
form hydrodynamic-type eddies. Such eddies, as in
hydrodynamic turbulence, decay in one eddy turnover time.

Below we consider the effect of 
imbalance on the turbulence decay time scale.
MHD turbulence can be described by opposite-traveling wave packets.
`Imbalance' means that wave packets traveling in one
direction
have significantly larger amplitudes than the other.
In astronomy, many energy sources are localized.
For example, SN explosions and OB winds are typical point energy sources.
Furthermore, astrophysical jets from YSOs are believed to be
highly collimated.
With these localized energy sources, it is natural to think
that interstellar turbulence is typically  imbalanced.

In this subsection, we explicitly relate the degree of imbalance and
the decay time scale of turbulence in the presence of a strong
uniform background field. 

\begin{figure}[t]
\epsscale{0.80}
\plotone{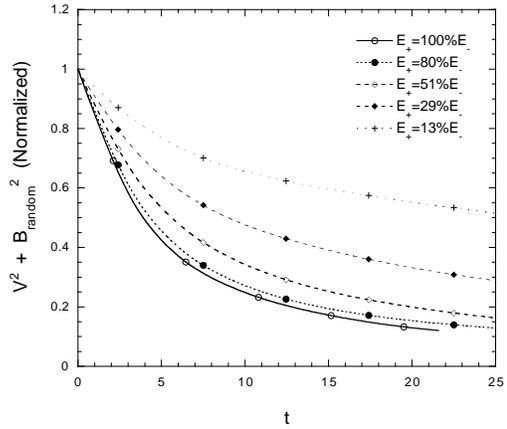}
\caption{  Decay of imbalanced turbulence.
                 Imbalanced cascade can extend
                decay time.
                Run 144H-$B_0$1.  (Cho et al. 2001)   
 }
\end{figure}

In Figure 2 we demonstrate that an imbalanced cascade does extend
the lifetime of MHD turbulence.
We use the run 144H-$B_0$1 to investigate the decay time scale.
We ran the simulation up to t=75 with non-zero driving forces.
Then at t=75, we turned off the driving forces
and let the turbulence decay.
At t=75, there is a slight imbalance between upward and downward moving
components ($E_+ = 0.499$ and $E_- = 0.40$).
This results from a natural fluctuation in the simulation.
The case of $(E_{+})_{t_0} = 80\% (E_{-})_{t_0}$ corresponds to
the simulation that starts off from this initial imbalance.
In other cases, we either increase or decrease the energy of ${\bf z}^{-}$
components and, by turning off the forcing terms, let the turbulence decay.
We can clearly observe that imbalanced turbulence extends
the decay time scale substantially.
Note that we normalized the initial energy to 1.
The y-axis is the total (=up $+$ down) energy.

In this section, we found that turbulence decay time can be slow.
If we consider the interstellar medium at larger scales,
it is definitely compressible, has a whole range of energy
injection/dissipation
scales (see Scalo 1987), and the relative role of vortical versus compressible
motions being unclear. Nevertheless, we believe that our
simplified
treatment may still elucidate some of the basic processes.
To what extend
this claim can be justified will be clear when we 
compare compressible and incompressible results. However, if we accept
that fast and slow MHD modes are subjected to fast collisionless damping
(see Minter \& Spangler 1997) the remaining modes are 
incompressible Alfven modes.  Those should
be well described by our model when turbulence is supersonic but
sub-Alfvenic. Our preliminary results (Cho \& Lazarian 2002) show
that the coupling of the modes is marginal even in compressible
regime.

\subsection{STRUCTURES BELOW VISCOUS CUT-OFF}

In hydrodynamic turbulence viscosity sets a minimal scale for
motion, with an exponential suppression of motion on smaller
scales.  Below the viscous cutoff the kinetic energy contained in a 
wavenumber band is 
dissipated at that scale, instead of being transferred to smaller scales.
This means the end of the hydrodynamic cascade, but in MHD turbulence
this is not the end of magnetic structure evolution.  For 
viscosity much larger than resistivity,
$\nu\gg\eta$, there will be a broad range of
scales where viscosity is important but resistivity is not.  On these
scales magnetic field structures will be created through a
combination of large scale shear and the small scale motions generated
by magnetic tension.  As a result, we expect
a power-law tail in the energy distribution, rather than an exponential
cutoff.  To our best knowledge, this is a completely new regime
of MHD turbulence. 
Ambipolar diffusion damps turbulent motions in ISM.
Therefore, the study of such a viscous damped MHD turbulence is
particularly important for many astrophysical processes below
the ambipolar diffusion scale in ISM.

In Cho, Lazarian, \& Vishniac (2002), we 
numerically demonstrate the existence of the power-law
magnetic energy spectrum below the viscous damping scale.
We use the same pseudo-spectral incompressible MHD code.
We use physical viscosity for velocity. The kinetic Reynolds number is
around 100. We use three different values of magnetic diffusion:
physical diffusion, 
hyper-diffusion with order 2, and hyper-diffusion with order 8.
We will present a theoretical model for this regime in an upcoming
paper (Lazarian, Vishniac, \& Cho 2002).

In Figure 3, we plot energy spectra.
The spectra consist of several parts.
{}First, the peak of the spectra corresponds to the energy injection scale.
Second, for $2<k<6$, kinetic and magnetic spectra follow a similar slope.
This part may be the short inertial range in general sense.
Third, magnetic and kinetic spectra decouple at $k\sim 6$.
Forth, for $20<k$, it appears that a new damped-scale inertial range emerges.
The emergence of the new inertial range is conspicuous in the order 8
hyperdiffusion simulation.

\begin{figure}[t]
\epsscale{0.80}
\plotone{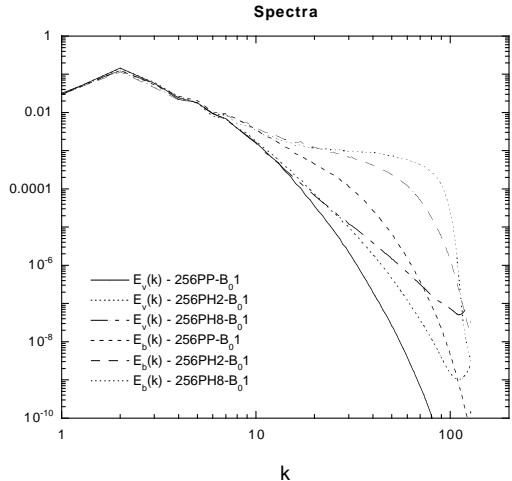}
\caption{  
   Energy spectra at t=15. From Cho et al. 2002.
   }
\end{figure}

\section{SUMMARY}

We described a pseudo-spectral MHD code and demonstrated its usefulness
in astrophysics through our recent works.
We showed that the rate at which MHD turbulence decays depends on the degree
of energy imbalance between wave packets traveling in opposite directions.
A substantial degree of imbalance can substantially 
extend the decay time scale of the MHD turbulence, which might be useful
for the study of star formation.
We also showed that magnetic fields can have rich structures below the
viscous cut-off scale, which implies that magnetic fields have stochastic
structures below the ambipolar diffusion scale in ISM and, therefore, affect
many astrophysical processes, such as
cosmic ray transports, magnetic reconnection, and thermal diffusion.
\acknowledgments{
   I thank Alex Lazarian, Ethan T. Vishniac, and Peter Goldreich for
valuable discussions. The introduction of this proceeding is partly
based on email communication with John Scalo.
This work was partially supported by National Computational Science
Alliance under CTS980010N and AST000010N and
utilized the NCSA SGI/CRAY Origin2000.
}

\end{document}